# LARGE-SCALE POWER-SPECTRUM FROM PECULIAR VELOCITIES

Tsafrir Kolatt[1,2] and Avishai Dekel[2]

[1]Harvard-Smithsonian Center for Astrophysics, 60 Garden st., Cambridge, MA 02138.
tsafrir@cfa.harvard.edu

[2]Racah Institute of Physics, The Hebrew University, Jerusalem 91904, Israel.
dekel@astro.huji.ac.il

## ABSTRACT

The power spectrum (PS) of *mass* density fluctuations, in the range $0.05 \leq k \leq 0.2\,h\,{\rm Mpc}^{-1}$, is derived from the Mark III catalog of peculiar velocities of galaxies, independent of "biasing". It is computed from the density field as recovered by POTENT with Gaussian smoothing of $12\,h^{-1}{\rm Mpc}$, within a sphere of radius $\sim 60\,h^{-1}{\rm Mpc}$ about the Local Group. The density is weighted inversely by the square of the errors. The PS is corrected for the effects of smoothing, random errors, sparse sampling and finite volume using mock catalogs that mimic in detail the Mark III catalog and the dynamics of our cosmological neighborhood. The mock catalogs are also used for error analysis.

The PS at $k=0.1\,h\,{\rm Mpc}^{-1}$ (for example) is $(4.6\pm1.4)\times 10^3\,\Omega^{-1.2}\,(h^{-1}{\rm Mpc})^3$, with a local logarithmic slope of $-1.45\pm0.5$. An itegration yields $\sigma_8\,\Omega^{0.6} \simeq 0.7-0.8$, depending on where the PS peak is. Direct comparisons of the mass PS with the galaxy PS derived from sky and redshift surveys show a similarity in shape and yield for $\beta \equiv \Omega^{0.6}/b$ values in the range $0.77 - 1.21$, with a typical error of $\pm 0.1$ per galaxy sample.

A comparison of the mass PS at $\sim 100\,h^{-1}{\rm Mpc}$ with the large-angle CMB fluctuations by COBE provides constraints on cosmological parameters and on the slope of the initial PS ($n$). The "standard" CDM model is marginally rejected at the $\sim 2\sigma$ level, while each of the following modifications leads to a good fit: $n \lesssim 1$, $\Omega_\nu \sim 0.3$, or $\Omega \lesssim 1$. Values of $\Omega$ as low as $\sim 0.2$ are ruled out with high confidence (independent of $\Lambda$).

*Subject headings:* cosmology: theory — cosmology: observation — dark matter — galaxies: distances and redshifts — large scale structure of universe

# 1. INTRODUCTION

The power spectrum of density fluctuations is the most common statistics used to quantify the large-scale structure in the universe (*e.g.* Peebles 1980; 1993). This statistics is useful for several reasons. If the initial fluctuations were a Gaussian random field, as commonly assumed, then the initial PS fully characterizes the statistical properties of the field, and it reflects the origin of fluctuations in the early universe. The PS is particularly useful because its development during the course of linear and quasi-linear evolution under gravitational instability (GI) obeys a simple, uniform growth rate, independent of scale. Then, the most pronounced effect of the dark matter in the plasma era is via a characteristic filtering of the initial PS on scales below $\sim 100$ $h^{-1}$Mpc, which makes the present PS on these scales reflect the type of dark matter that dominates the mass in the universe. Furthermore, the elegant mathematical properties of Fourier analysis make the PS relatively straightforward to compute, in k-space or via its Fourier transform, the spatial two-point correlation function. The PS in a certain range of scales can be determined by the data better than most other common statistics for large-scale structure.

The power spectra of the distribution of galaxies and clusters have been computed from many different samples, in two dimensions on the sky, or in three dimensions, corrected from redshift space (see §4). New determinations with smaller errors are expected from very large redshift surveys in progress (2DF, SDSS). However, these power spectra correspond to objects which are not necessarily unbiased tracers of the underlying mass distribution that is directly related to theory (*e.g.* Dekel & Rees 1987). Clear evidence for this bias is provided by the fact that galaxies and clusters of different types are observed to cluster differently (*e.g.* Dressler 1980). It would therefore be naive to assume that any of the galaxy power-spectra directly reflects the mass PS. Instead, one wishes to measure the mass PS from dynamical data, bypassing the complex issues involved in galaxy formation, so called galaxy "biasing". Such dynamical information can be provided by peculiar velocities, by gravitational lensing effects, and by fluctuations in the cosmic microwave background (CMB).

The accumulating catalogs of peculiar velocities of galaxies enable a direct determination of the mass PS under the natural assumption that the galaxies are honest tracers of the large-scale velocity field induced by gravity. Here, we make a first attempt to determine this mass PS in the wavelength range $30 - 120$ $h^{-1}$Mpc. We do it using the smoothed density field recovered by the POTENT procedure from the *Mark III Catalog of Peculiar Velocities* (see §2). An alternative determination of the mass PS from the raw Mark III data, using likelihood analysis to constrain the parameters of a linear, Gaussian prior model for the PS, is provided in an associated paper (Zaroubi *et al.* 1996).

A direct comparison of the mass PS derived here with the various galaxy power spectra is used to estimate the parameter(s) $\beta_x \equiv \Omega^{0.6}/b_x$, where $\Omega$ is the universal density parameter, and $b_x$ is the linear biasing parameter for the specific type of galaxies in the given sample.

Then, a comparison of the PS derived from velocities on scales $\lesssim 100$ $h^{-1}$Mpc with the dynamical fluctuations observed in the CMB by COBE over angles corresponding to comoving $\gtrsim 1000$ $h^{-1}$Mpc at $z \sim 10^3$ provides a leverage for constraining the cosmological parameters ($\Omega$, $H_0$) together with the linear fluctuation power index ($n$) and the dark matter species.

In §2 we give a brief description of the Mark III catalog and the POTENT reconstruction. We then describe in §3 the method for computing the raw PS and the corrections applied using mock catalogs in order to recover the true underlying PS and to estimate the associated errors. In §4 we compare the mass density PS from peculiar velocities with galaxy power spectra from galaxy density surveys and obtain estimates of $\beta$. In §5 we use the mass-density PS and the COBE measurements to constrain cosmological models. We discuss our results and summarize our conclusions in §6.

## 2. DATA AND POTENT ANALYSIS

### 2.1 The data

The Mark III Catalog of Peculiar Velocities (Willick *et al.* 1995 WI; 1996a WII; 1996b WIII) consists of more than 3000 galaxies from several different data sets of spiral and elliptical/S0 galaxies with distances inferred by the forward Tully-Fisher and $D_n-\sigma$ methods, which were re-calibrated and self-consistently put together as a homogeneous catalog for velocity analysis. The cluster data sets are treated in WI. The field galaxies are calibrated and grouped in order to minimize Malmquist biases in WII. The final catalog is tabulated in WIII.

The catalog provides radial peculiar velocities and inferred distances properly corrected for inhomogeneous Malmquist bias for $\sim 1200$ objects, ranging from isolated field galaxies to rich clusters. The associated errors are on the order of $17-21\%$ of the distance per galaxy. These data enable a reasonable recovery of the dynamical fields with Gaussian smoothing of radius $\sim 12$ h$^{-1}$Mpc out to a distance of $\sim 60$ h$^{-1}$Mpc from the Local Group.

### 2.2 POTENT reconstruction

The POTENT method recovers the smoothed dynamical fluctuation fields of potential, velocity and mass density from the observed radial peculiar velocities, under quasi-linear GI (Dekel *et al.* 1990; Bertschinger *et al.* 1990; Nusser *et al.* 1991; Dekel *et al.* 1996). Given the sparsely-sampled radial velocities, POTENT first computes a *smoothed* radial-velocity field $u(\boldsymbol{x})$ in a spherical grid, using a linear tensor window function which mimics a Gaussian of radius 12 h$^{-1}$Mpc (G12). Weighting inversely by the local density near each object mimics equal-volume averaging which minimizes the bias due to sampling gradients. Weighting inversely by the distance variance of each object, $\sigma_i^2$, reduces the random effects of the distance errors.

The velocity field is recovered under the assumption of potential flow: $\boldsymbol{v}(\boldsymbol{x}) = -\boldsymbol{\nabla}\Phi(\boldsymbol{x})$. According to linear theory, any vorticity mode decays in time as the universe expands, and based on Kelvin's circulation theorem the flow remains vorticity-free in the mildly-nonlinear regime as long as the flow is laminar. This has been shown to be a good approximation when collapsed regions are properly smoothed over. The velocity potential can thus be calculated by integrating the radial velocity along radial rays, $\Phi(\boldsymbol{x}) = -\int_0^r u(r', \theta, \phi) dr'$. Differentiating $\Phi$ in the transverse directions recovers the two missing velocity components.

The underlying mass-density fluctuation field, $\delta(\boldsymbol{x})$, is then computed by

$$\delta(\boldsymbol{x}) = \left\| I - f(\Omega)^{-1} \frac{\partial \boldsymbol{v}}{\partial \boldsymbol{x}} \right\| - 1 \;, \tag{1}$$

where the bars denote the Jacobian determinant and $I$ is the unit matrix. In our notation the Hubble constant is set to unity and distances are measured in $\mathrm{km\,s^{-1}}$. Then $f(\Omega) \equiv \dot{D}/D \simeq \Omega^{0.6}$ where $D(t)$ is the linear growth factor (Peebles 1980). Eq. 1 is the solution to the continuity equation under the Zel'dovich assumption that particle displacements evolve in a universal rate (Zel'dovich 1970; Nusser *et al.* 1991). This nonlinear approximation, which reduces in the linear regime to the familiar $\delta = -f(\Omega)^{-1}\boldsymbol{\nabla}\cdot\boldsymbol{v}$, approximates the true density in N-body simulations with an *rms* error less than 0.1 over the range $-0.8 \leq \delta \leq 4.5$ (Mancinelli *et al.* 1994).

The largest source of random uncertainty are the *distance errors*, $\sigma_i$, of the individual galaxies. The final errors are assessed by Monte-Carlo simulations, where the input distances are perturbed at random using a Gaussian of standard deviation $\sigma_i$, and each artificial sample is fed into POTENT. The standard deviation of the recovered $\delta$ at each grid point over the Monte-Carlo simulations, $\sigma_\delta(\boldsymbol{x})$, serves as our estimate for the random distance error. In the well-sampled regions (out to 40 $\mathrm{h^{-1}Mpc}$ in most directions and beyond 60 $\mathrm{h^{-1}Mpc}$ in certain directions) the measurement errors are below 0.3, but they exceed unity in certain poorly sampled, noisy regions at large distances.

Our reconstructed fields are also subject to systematic errors: the inhomogeneous Malmquist bias (IMB) and a sampling gradient bias. The IMB is due to the coupling of distance errors and the clumpy distribution of galaxies from which the sample has been selected. Most of the IMB has been removed from the raw data by heavy grouping in redshift space and by correcting the estimated distances using the IRAS 1.2Jy density field (A. Yahil & M. Strauss, private communication, based on Fisher *et al.* 1995) as a tracer of the underlying galaxy density from which the Mark III data sets were selected (Willick 1991; 1994; Dekel 1994; Dekel *et al.* 1996). The IMB correction in $\delta$ is typically limited to $\sim 10-20\%$ reduction in density contrast in the peaks, which is small compared to the random errors at distances typical to the Mark III data.

## 3. COMPUTING THE POWER SPECTRUM

### 3.1 Observed Power Spectrum

The POTENT output provides the G12-smoothed density fluctuation field, $\delta(\boldsymbol{r})$, at the points of a cubic grid of 5 $\mathrm{h^{-1}Mpc}$ spacing within a sphere of radius 80 $\mathrm{h^{-1}Mpc}$, and the associated random error field $\sigma_\delta(\boldsymbol{x})$.

The density field is weighted by $\sigma_\delta^{-2}(\boldsymbol{x})$ in order to minimize the effect of random errors. It is then zero-padded beyond a radius of 80 $\mathrm{h^{-1}Mpc}$ to fill a cubic, periodic box of side 160 $\mathrm{h^{-1}Mpc}$, and volume $V$. This discrete density field is Fourier transformed using FFT, with the convention

$$\delta(\boldsymbol{x}) = V^{-1/2} \sum \tilde{\delta}(\boldsymbol{k})\, e^{i\boldsymbol{k}\cdot\boldsymbol{x}}, \quad \tilde{\delta}(\boldsymbol{k}) = V^{-1/2} \int \delta(\boldsymbol{x})\, e^{-i\boldsymbol{k}\cdot\boldsymbol{x}} \mathrm{d}\boldsymbol{x}, \quad P(k) = \langle|\tilde{\delta}(\boldsymbol{k})|^2\rangle, \quad (2)$$

and a $k$-grid of spacing $2\pi/160$ $\mathrm{h\,Mpc^{-1}}$. The power spectrum is computed in five equal logarithmic bins of width $\Delta \log k = 0.24$, starting from $k = 0.039\,\mathrm{h\,Mpc^{-1}}$ (the smallest wavenumber). In each bin we compute $P(k)$ by the mean square amplitude of the Fourier transform over the wavevectors that lie in that bin. The resultant $P(k)$ is assigned to a bin center that is defined by the average wavenumber $k$ over these wavevectors. We end up with $P(k)$ estimates in five bins centered on $k = 0.06, 0.10, 0.17, 0.30, 0.52\,h\,\mathrm{Mpc}^{-1}$. For reasons explained at the end of §3.1, our statistical estimates are based only on the first three bins. We term the product of the above procedure the "observed" PS, $P_o(k)$. The next task is to recover the true PS out of this biased and noisy measurement.

### 3.2 Recovering the True Power Spectrum

There is a long way between the true fluctuation signal and the observed PS. The data have been contaminated by distance errors, and sampled sparsely and non-uniformly. They have then been smoothed heavily and subjected to the POTENT analysis where quasi-linear approximations were used to derive the density field and the associated errors on a uniform grid. The density field has been taken from within a finite volume, inversely weighted by the errors, and zero-padded before an FFT procedure has been applied and the PS has been computed in finite bins. Systematic biases were introduced in every step of this procedure, which are hard to correct for by standard analytic tools. In particular, an approximate deconvolution procedure, similar to the one used by Park *et al.* (1994), fails because the Fourier transform of the spatial sampling window is not properly peaked about $k = 0$. We therefore chose to apply an empirical correction procedure based on carefully-designed mock catalogs.

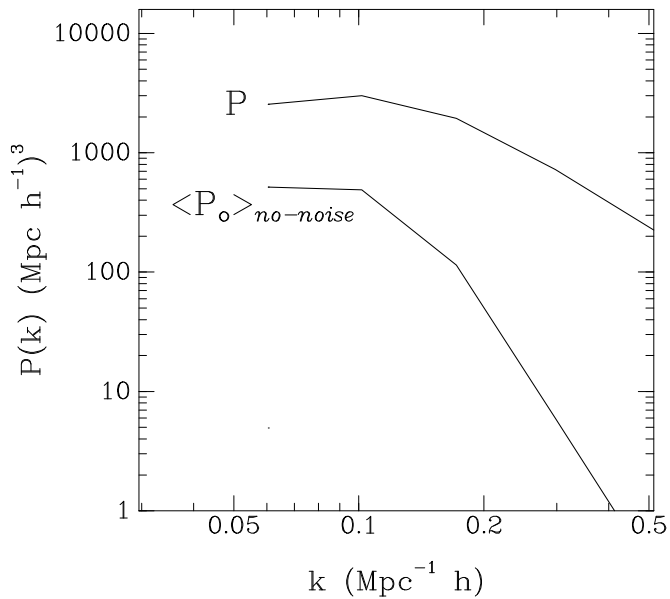

**Figure 1:** The true PS of the mass density in the simulation from which the mock catalogs were drawn, $P(k)$, and the average raw PS as computed from the G12 POTENT reconstructions of the ensemble of unperturbed mock catalogs, $\langle P_o(k) \rangle_{no-noise}$. The correction function $M^{-1}(k)$ is the ratio between the two.

The systematic errors introduced by the above procedure are modeled by a simple model:
$$P_o(k) = M(k)[P(k) + N(k)], \qquad (3)$$

where $P(k)$ is the true signal, $N(k)$ is the PS of the noise, and $M(k)$ represents the effects of sampling, smoothing, applying a spatial window etc. The correction functions $M(k)$ and $N(k)$ are to be derived from Monte Carlo mock catalogs based on simulations that mimic in detail the underlying structure of the real universe and the Mark III catalog (Kolatt *et al.* 1996).

It is not obvious *a priori* that the simple model of Eq. 3 is adequate. One could imagine, for example, that the observed $P_o(k)$ is a convolution of the true $P(k)$ with some filter that mixes different scales. In this case $M(k)$ would be a tensor. We adopt the simple model based on it's reasonable success in recovering the PS in the mock catalogs, as demonstrated below.

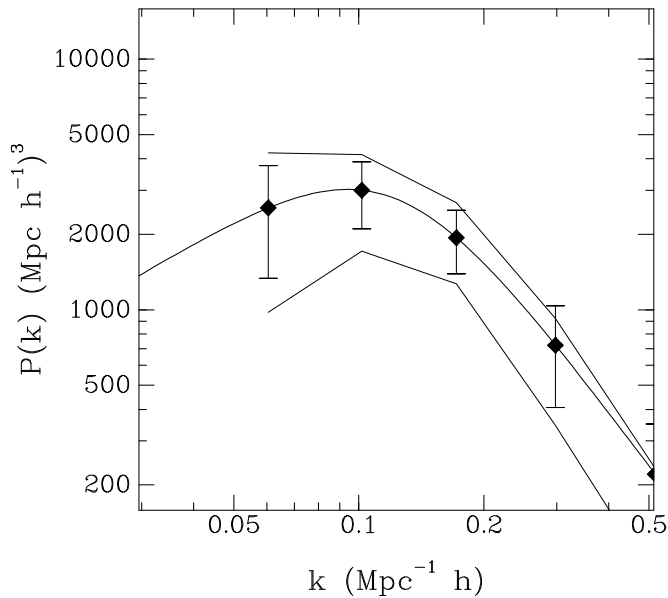

**Figure 2a:** The errors in the recovered PS, derived as the standard deviation in each bin over the mock catalogs. Shown also are the two worst cases among the 20 mock catalogs (thin curves), in comparison with the true PS of the simulation (thick curve).

The factor $M(k)$ is derived first from mock catalogs that were sparsely sampled but were *not* perturbed by distance errors, via

$$M(k)^{-1} = P(k)/\langle P_o(k)\rangle_{no-noise}, \qquad (4)$$

where $P(k)$ here is the known signal in the simulations, and the averaging is over twenty unperturbed mock catalogs. Figure 1 shows $P(k)$ and $\langle P_o(k)\rangle_{no-noise}$ as a function of $k$. The correction factor $M^{-1}$ varies from $\simeq 5$ at $k=0.06$ to $\simeq 30$ at $k=0.2$, and to above 100 beyond $k=0.3$. One source of this variation is the G12 smoothing.

Then, the noise PS, $N(k)$, is computed by

$$N(k) = M(k)^{-1}\langle P_o(k)\rangle_{noise} - P(k), \qquad (5)$$

where the averaging this time is over fully-perturbed mock catalogs.

Equipped with the correction functions $M(k)$ and $N(k)$, the $P_o(k)$ observed from the real universe is corrected to yield the true PS by

$$P(k) = M(k)^{-1}P_o(k) - N(k). \qquad (6)$$

To test this correction procedure, we apply it first to the PS observed from each of twenty perturbed mock catalogs. The corrected power spectra can be compared to the known true PS of the matter in the simulation. Figure 2a shows the two worst cases, highest and lowest, out of the twenty corrected power spectra.

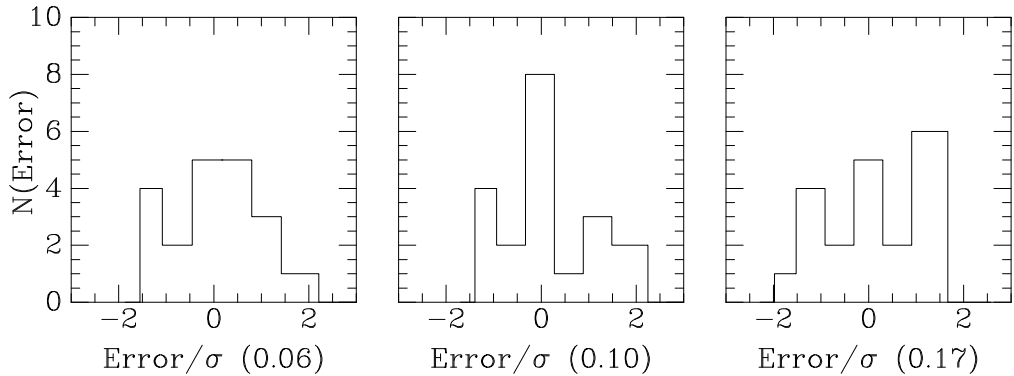

**Figure 2b:** The error distribution for 20 mock catalogs in the three first bins. The errors are measured in units of the standard deviation for each bin, and the wavenumber for the bin center (in $h\,\mathrm{Mpc}^{-1}$) is given in brackets.

Note that for each individual realization the values of the PS in the different bins are not entirely independent. For most cases these values are all either above the true PS or below it over a wide range of wavenumbers. In order to evaluate the degree of independence between the bins, we computed the ratios of the off-diagonal terms to the diagonal terms in the covariance matrix. These ratios were averaged over the 20 mock catalogs. The worst ratio, of 0.30, is obtained for the first two adjacent bins, 1 and 2. Bins 2 and 3 show an average ratio of 0.24, and the correlation between bins 1 and 3 reduces to a ratio of 0.18.

The error bars in Fig. 2a correspond to the standard deviation over the corrected power spectra from the twenty perturbed mock catalogs. The actual distributions of the 20 perturbed power spectra about the true values are shown bin by bin in Figure 2b. The distributions crudely resemble a Gaussian distribution only within the inner $\pm 1\sigma$, and they indicate excessive tails with some tendency for positive skewness.

In principle, the effects of sampling and noise on the PS may depend on the PS signal itself, which would require a signal-dependent correction procedure. In order to test for such dependence, we have generated a second set of mock catalogs with higher true PS by running the N-body simulation forward in time from $\sigma_8 = 0.7$ to $\sigma_8 = 1.14$. Figure 3 shows the average of the PS from the low-amplitude set of catalogs, corrected by $M(k)$ and $N(k)$ which were derived from the high-amplitude set, and vice versa. We find that the correction procedure is robust with respect to the amplitude of the true PS, which means that at least as long as the true PS in the universe is between the two simulated cases, we should be able to use either set of mock catalogs for deriving the correction functions. The range of wave-numbers where the reconstruction is "reliable" can now be determined, for example, by the requirement that for each of the 20 perturbed data sets (*i.e.* with 95% probability) the correction procedure, using either set of correction functions, always yields $P(k) > 0$. This range is found to be $k \leq 0.2$. We hence restrict any statistical analysis to the three bins in this $k$ range.

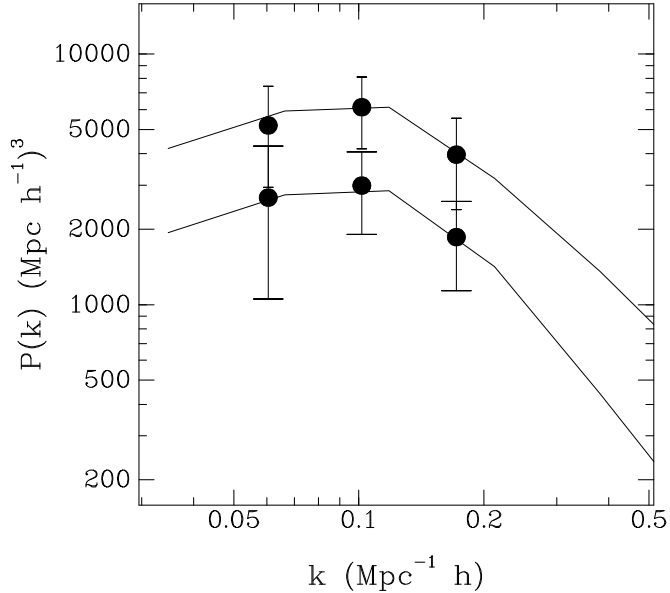

**Figure 3:** The power spectra of the mock catalogs from each of the two simulations ($\sigma_8 = 0.7$ and $1.14$), corrected by correction functions derived from the other simulation. Shown in each bin are the average (symbols) and standard deviation over the mock catalogs of each simulation. The curves are the corresponding true power spectra in the simulations. This demonstrates that the signal dependence of the correction procedure is negligible.

Figure 4 shows the PS as derived from the real peculiar velocity data, corrected alternatively by the low-amplitude mock catalogs or the high-amplitude set. The two corrections yield similar results.

### 3.3 The Resultant Power Spectrum

Figure 5 finally shows our best estimate of the *true* mass power spectrum, which we take to be the average of the two curves of Fig. 4. Table 1 contains the same information for the full range of five $k$ bins. Note that since the raw data is peculiar velocities, the mass PS is only determined up to a multiplicative factor, $f^2(\Omega) \approx \Omega^{1.2}$. The $1\sigma$ error in each bin ($\sigma_v$) is the product of the average signal and the average noise-to-signal ratio in the two sets of mock catalogs. Recall that despite the fact that we have used broad bins, the PS values in the different bins are not fully independent of each other.

The resultant mass PS can be characterized, for example, by its amplitude at $k = 0.1\,h\,\mathrm{Mpc}^{-1}$, $P_{0.1}f^2 = (4.6 \pm 1.4) \times 10^3\,(h^{-1}\mathrm{Mpc})^3$, and the local logarithmic slope of

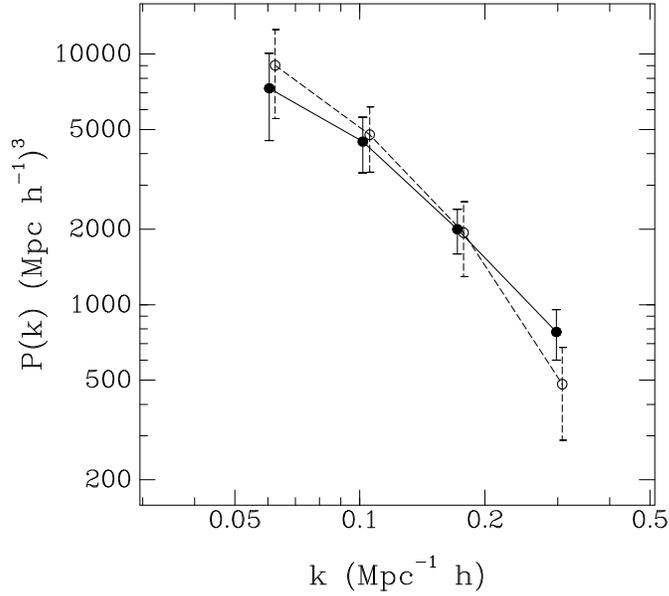

**Figure 4:** The PS of the real data as recovered using alternatively the correction functions that were derived from the two different simulations ($\sigma_8 = 0.7$ solid, $\sigma = 1.14$ dashed and slightly shifted to the right). The errors are as in Fig. 2.

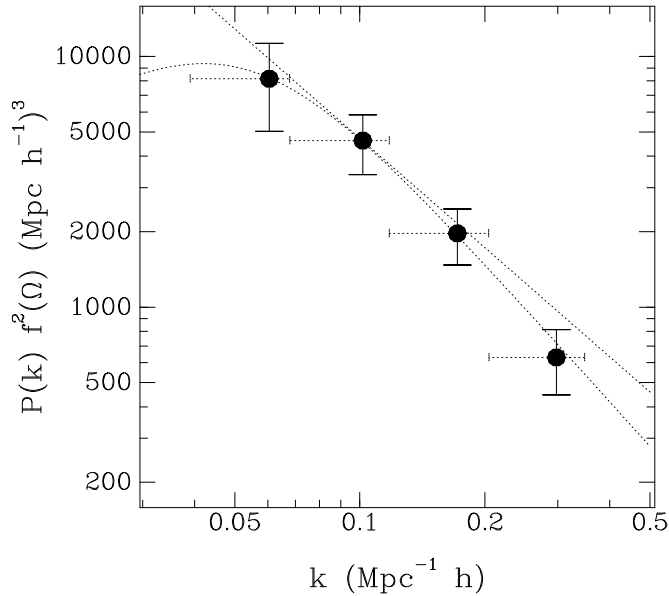

**Figure 5:** The mass density PS as computed from the Mark III peculiar velocities. The errors, based on the mock catalogs, indicate measurement errors about the PS in our local neighborhood (not including cosmic scatter). The horizontal dotted lines mark the wavenumber bins. The dotted curves, shown for reference, are (a) the functional fit of Eq. 7 with $k_c = 0.05$, $k_0 = 0.30$, $m = 1.84$, and (b) a straight line of slope $-1.45$ tangent to the PS at $k = 0.1$

$-m_{0.1} = -1.45 \pm \sim 0.50$ (a dotted line in Fig.5). The error in the slope is estimated by an

| $k$ | $f^2(\Omega)P(k)$ |
|---|---|
| 0.061 | 8157 $\pm$ 3127 |
| 0.102 | 4620 $\pm$ 1240 |
| 0.172 | 1968 $\pm$ 495 |
| 0.297 | 629 $\pm$ 183 |
| 0.517 | 108 $\pm$ 33 |

**Table 1:** The PS values (in $(h^{-1}\mathrm{Mpc})^3$) in all five bins centered on the wavenumber $k$ (in $h\,\mathrm{Mpc}^{-1}$). Only the first three bins are used for statistical purposes for reasons explained in §3.2.

eye-ball fit using the two disjoint data points and errors at $k = 0.06$ and $0.17$ under the assumption that they are independent. A similar scatter is obtained for $m$ of Eq. 7 below when fitting it to each of the 20 mock catalogs (varying $m$ and $k_0$ while $k_c$ is fixed at it's true value).

A useful dimensionless function corresponding to $P(k)$ is $\Delta^2(k) = k^3/(2\pi^2)P(k)$, which is the local contribution to the density variance from a unit logarithmic interval in $k$, and is roughly equal to $\langle(\delta M/M)^2\rangle$ on a scale corresponding to $k$. Our result for $P_{0.1}$ thus corresponds to $\Delta^2_{0.1}f^2 = 0.233 \pm 0.071$. One can also see a slight tendency for flattening of the PS towards a possible peak beyond the longest wavelength bin, but this peak may probably be better constrained by information on larger scales, such as the bulk velocity over the sampled volume.

It may also be useful to use a functional fit of the form

$$P(k)f^2 = \frac{2\pi^2}{k^3}\frac{(k/k_0)^{3-m}}{1 + (k/k_c)^{-(m+n)}}. \tag{7}$$

The parameters are two inverse length scales, associated with the peak ($k_c$) and the scale of nonlinearity ($k_0$), and a logarithmic slope ($-m$) in the small-scale regime. (Note: $m = 3-\gamma$, where $\gamma$ is the logarithmic slope of the two-point correlation function.) The asymptotic slope on large scales ($n$) is hardly constrained by the velocity data alone, and is assumed here to be $n = 1$ without affecting the fit. The fit yields $m = 1.85 \pm 0.20$ and $k_0 = 0.30 \pm 0.05$, with $k_c$ weakly constrained to be $\sim 0.05$ or smaller (a dotted line in Fig. 5). The errors quoted here for each parameter are obtained by fixing all the other parameters at their best values and assuming that the data points are independent of one another (see discussion in §5.4).

The corresponding standard deviation of $\delta$ in a top-hat sphere of radius 8 $h^{-1}$Mpc is $\sigma_8 f = 0.71 - 0.77\,(\pm 0.12)$ for $k_c = 0.06 - 0.03$ respectively.

## 4. GALAXIES VERSUS MASS – THE $\beta$ PARAMETERS

The power spectra of the galaxy distribution over the relevant range of scales have been measured from several different surveys of galaxies, either angular on the sky or three-dimensional in redshift space. The relation between these galaxy PS and the mass PS derived from velocities depends on $\Omega$ and on how galaxies trace mass. If we define a linear biasing factor for galaxy type "$x$" by the effective ratio $b_x^2 = P_x(k)/P(k)$ in the vicinity of $k \sim 0.1\,\mathrm{h\,Mpc^{-1}}$, linear GI theory enables a straightforward determination of the corresponding parameters $\beta_x = f(\Omega)/b_x$.

The derivation of a real-space $P(k)$ from a measured redshift-space $P^s(k)$ involves a correction for redshift distortions, which can be crudely approximated by $P_x^s(k) = (1 + 2\beta_x/3 + \beta_x^2/5)\,P_x(k)$ (Kaiser 1987). Therefore, the ratio of the observables, $P_v(k) = P(k)f^2$ from peculiar velocities and $P_x^s(k)$ from redshifts, is a simple function of $\beta_x$,

$$\frac{P_v(k)}{P_x^s(k)} = F(\beta_x) \equiv \frac{\beta_x^2}{1 + 2\beta_x/3 + \beta_x^2/5}. \qquad (8)$$

In order to best exploit the data we estimate the average $\beta_x$ in the available range of scales by minimizing the sum of residuals over the bins,

$$\chi^2 = \sum_{i=1}^{3} \frac{[P_v(k_i) - F(\beta_x)P_x^s(k_i)]^2}{\sigma_v^2(k_i) + F^2(\beta_x)\sigma_x^2(k_i)}\,, \qquad (9)$$

where $\sigma_v$ is the measurement error in $P_v$.

We consider the galaxy power spectra derived from the following galaxy surveys:

1) IRAS 1.2Jy (Fisher *et al.* 1993; Yahil, private communication),
2) IRAS QDOT 1 in 6 (Feldman *et al.* 1994),
3) CfA2 and SSRS2 (Park *et al.* 1994; Vogeley *et al.* 1992; da Costa *et al.* 1994),
4) Las Campanas LCRS (Lin 1995),
5) APM (Baugh & Efstathiou 1993; Tadros & Efstathiou 1995).

For CfA2+SSRS2 we use the PS as computed by the authors in redshift space, within their box of 130 h$^{-1}$Mpc (Vogeley, private communication). For the other surveys we use functional fits by the authors, translated to the form of Eq. 7. For the angular APM survey, the fit is independent of redshift distortions. For IRAS 1.2Jy and QDOT, the fits were corrected by the authors for redshift distortions assuming $\beta = 1$. For LCRS the original fit is provided in redshift space. The fit parameters are listed in Table 2, where $k_0$ (original) is the authors quoted $k_0$.

For the likelihood analysis we assume, quite crudely, that the errors of the galaxy PS in the three points used are of 10% (the errors at smaller wavenumbers get much bigger), and that the measurements in these three points are independent of each other. These errors mostly come from uncertainties in the deconvolution of the sampling window (*e.g.* Park *et al.* 1992) and in the correction for redshift distortions (*e.g.* Zaroubi & Hoffman 1995). Note that the cosmic scatter in this analysis is reduced compared to the application in the next section because the sampling volumes of the galaxy surveys have significant overlaps with the volume sampled by the peculiar velocities.

The values of $\beta$ and their $1\sigma$ errors as obtained by the likelihood analysis for the different galaxy samples are listed in Table 2. The best fits span the range $0.77 - 1.21$,

| Survey | $m$ | $k_c$ | $k_0$ (original) | $k_0$ | $\beta$ |
|---|---|---|---|---|---|
| CfA2(130)+SSRS2 | – | – | – | – | $0.77 \pm 0.11$ |
| APM | 1.4 | 0.020 | 0.19 | 0.19 | $0.80 \pm 0.10$ |
| QDOT 1:6 | 1.4 | 0.033 | 0.24 ($r-space$, $\beta=1$) | 0.24 | $0.96 \pm 0.08$ |
| LCRS | 1.8 | 0.060 | 0.16 ($z-space$) | 0.28 | $0.99 \pm 0.13$ |
| IRAS 1.2 | 1.34 | 0.045 | 0.29 ($r-space$, $\beta=1$) | 0.31 | $1.21 \pm 0.10$ |

**Table 2:** The fit parameters (Eq. 7) for several galaxy surveys ($k$ in units of $h\,\mathrm{Mpc}^{-1}$) and the $k_0$ and $\beta$ values obtained by the maximum likelihood analysis. with typical error of 0.1. Also listed are the values of $k_0$ in r-space, corrected for redshift distortions using the corresponding best-fit values of $\beta$.

Figure 6 shows the galaxy power spectra, all corrected for redshift distortions using Kaiser's approximation with the appropriate best-fit value of $\beta$ ($k_0$) from Table 2. The mass PS is shown in comparison for $\Omega = 1$ (solid circles). The values of $\beta_x^2$ (for any $\Omega$) can be read directly from this figure as the effective ratios between the power spectra of mass (with $\Omega = 1$) and galaxies. For $\Omega = 1$, the biasing parameters are all of order unity, with the optical galaxies of APM and CfA2+SSRS slightly biased ($b_x = 1.25 - 1.3$), IRAS 1.2Jy galaxies slightly anti-biased ($b_x = 0.83$), and the galaxies of QDOT and LCRS unbiased.

The figure also displays the mass PS for $\Omega = 0.3$ (open circles), showing the corresponding $b_x^2$ via the ratios of galaxy to mass PS. In this case, the galaxies are all severely anti-biased, with biasing parameters of $0.4 - 0.6$. It would be hard to imagine a scenario of galaxy formation that could produce antibiasing at this level for all galaxy types.

A similar conclusion is obtained from a different point of view, by comparing alternatively the *rms* fluctuations $\sigma_8$ of galaxies and mass. We found for the mass (§3) $\sigma_8 f = 0.71 - 0.77$. If $f = 1$, then the value of $\sigma_8$ lies reasonably close and between the corresponding values observed for optical galaxies ($\simeq 0.95$) and for IRAS 1.2Jy galaxies ($\simeq 0.6$) (*e.g.* de Lapparent, Geller, & Huchra 1988; Strauss *et al.* 1996). If $f \simeq 0.5$ say, then $\sigma_8 \simeq 1.5$ and all the galaxies are severely anti-biased.

The logarithmic slopes of all the power spectra in the range of comparison are similar to one another, $m_{0.1} \approx -1.4$ at $k = 0.1$, showing no significant evidence for scale dependence in the biasing within this limited range of scales. The turnover could occur, within the errors, anywhere below $k = 0.06\,h\,\mathrm{Mpc}^{-1}$, thus providing no strong constraint on the cosmological parameters.

Power spectra were also derived for the distribution of radio galaxies (Peacock & Nicholson 1991), for rich clusters of galaxies in the Abell/ACO catalog (Peacock & West 1992) and for clusters in the APM survey (Dalton *et al.* 1994). The PS for the $z < 0.1$ sample of radio galaxies is best fit in redshift space by Eq. (7) with parameters similar to those found for the APM galaxies except that the amplitude is 3.3 times larger (*i.e.* $k_0 = 0.09$). After correcting for redshift distortions using Eq. (8), the comparison with the mass PS yields $\beta_{radio} \simeq 0.50$. The PS of the Abell/ACO $R \geq 1$ clusters has been fit in redshift space by Eq. (7) with $m = 1.4$, $k_0 = 0.048$, and $k_c = 0.025$. The comparison with the mass PS yields $\beta_{R \geq 1} \simeq 0.31$. The APM clusters, which are on average less rich, have a PS amplitude lower by a factor of $\sim 1.5$, and therefore $\beta_{APMcl} \simeq 0.46$. These low values of $\beta$ for clusters are consistent with other rough estimates (*e.g.* Plionis 1995), confirming the fact that the clusters are severely-biased tracers of the mass. The relatively high biasing parameter for radio galaxies can probably be attributed to their known correlation with clusters of galaxies.

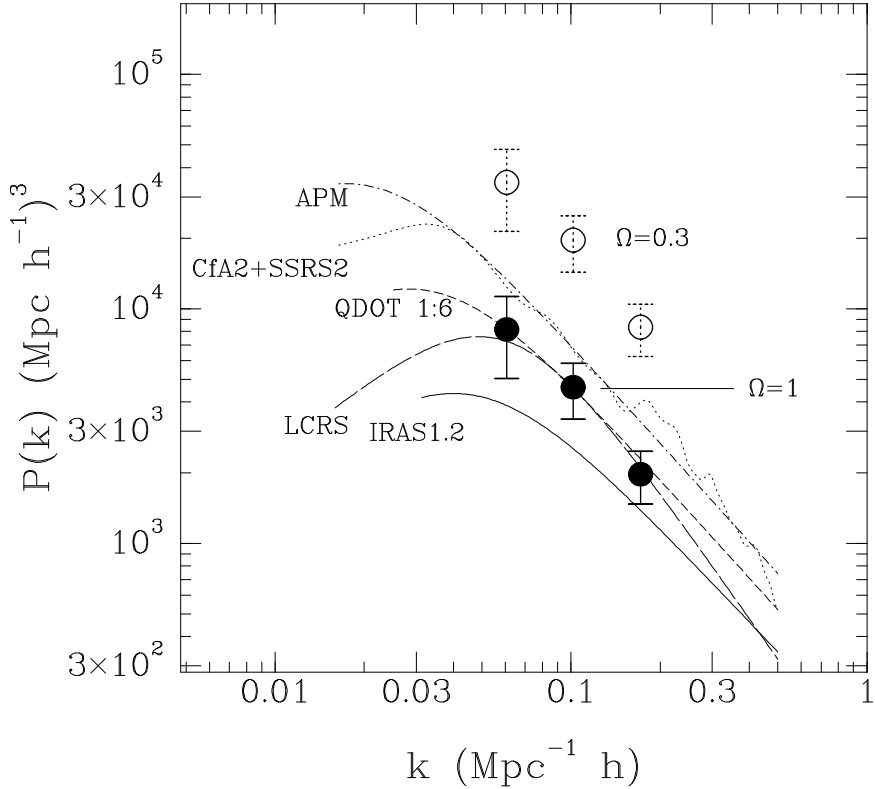

**Figure 6:** Power spectrum estimates from various samples, all in real space, corrected from red-shift space by the best value $\beta$ (table 1). The peculiar velocities power spectrum is marked by solid dots ($\Omega = 1$) and open dots ($\Omega = 0.3$). The CfA galaxies line is drawn by interpolation between data points as published by the authors. All other lines are drawn from the fitting formulae as calculated by the authors. Lines span according to each survey detection range.

## 5. COMPARISON WITH COBE AND THEORETICAL MODELS

The observed peculiar velocities constrain the amplitude of dynamical fluctuations on scales $\lesssim 100$ h$^{-1}$Mpc, while the fluctuations observed in the CMB by COBE's DMR experiment span an angle range corresponding to $\gtrsim 1000$ h$^{-1}$Mpc, comoving, at $z \sim 10^3$. The relation between the quantities measured by these data involves the spatial shape of the PS and its time evolution, which depend on the cosmological parameters ($\Omega$, $\Lambda$, $\Omega_b$, $h$), the shape of the initial PS which is still valid on large scales (power index $n$), and to some extent the small-scale filtering, which also depends on the hot dark matter content ($\Omega_\nu$). The wide range of scales between COBE's measurements and the peculiar velocity data can be used as an effective leverage for constraining these parameters.

A similar comparison to COBE could be done by using the PS from galaxy *density* surveys (Fisher *et al.* 1993, Park *et al.* 1994, Lin 1995, Vogeley 1995), where the measurement errors are smaller than in the case of peculiar velocities and the PS extends to larger scales (sometimes beyond the PS peak), but this comparison is contaminated by unknown, sample dependent, galaxy biasing and by redshift distortions, while the current comparison is free of these effects.

The use of the mass PS over the whole range where it has been measured is advantageous over using the large-scale bulk velocity alone, because the PS spans locally (in $k$) the range of wavenumbers on the short-wavelength side of the peak ($k > k_c$), where the PS is steep and therefore sensitive to a horizontal shift in scale. Such a shift is commonly associated with the value of the product $\Omega h$, reflecting the horizon scale when the universe turned matter dominated. This steepness promises that the PS in this range can provide a sensitive measure of the value of $\Omega h$. We restrict ourself to specific theoretical models, where the PS is specified at $z \sim 10^3$ and is assumed to grow according to GI since then. Over the scales where the mass PS has been derived, and on larger scales, we adopt the linear approximation, where the PS is assumed to evolve according to a universal growth rate on all scales (except for the case of hot dark matter – see below). The justification for this approximation is demonstrated in Figure 7, which compares the linear and non-linear power spectra on the relevant scales. The continuous line in Fig. 7 depicts the linear CDM power spectrum normalized to $\sigma_8 = 0.77$ and smoothed G12. This power spectrum served to generate the random initial conditions for 20 N-body simulations. Each of the 20 realizations was dynamically evolved using a PM code (Bertschinger & Gelb 1991) until it reached $\sigma_8 = 0.77$. It was then smoothed G12 and a PS was computed. The points in figure 7 represent the averages of the PS over the 20 simulations and the errors are the standard deviation among this ensemble of simulations. There is a pleasant agreement between the theoretical curve and the evolved smoothed PS results. The validity of the linear growth rate even in the weakly-nonlinear regime can be attributed to the compensation of the faster nonlinear growth in clusters by the slower nonlinear growth in voids. (*e.g.* Nusser & Dekel 1992, Hamilton & Taylor 1996).

### 5.1 The Model Power Spectra

In most cases we model the theoretical power spectra by a function of the form

$$P(k) = AT^2(k)k^n, \qquad (10)$$

where $A$ is a normalization constant that is not predicted by theory, and $T(k)$ is the transfer function which, together with $n$, characterizes the model. The power index, $n$, reflects the initial fluctuations. Most theories of the origin of fluctuations, by Inflation or by topological defects, predict that it should be of order unity, or slightly smaller (*e.g.* Steinhardt 1995).

We start with the following families of CDM models (with varying values of $h$):

1) sCDM, the "standard" CDM model with $\Omega = 1$, $\Lambda = 0$ and $n = 1$.
2) TCDM, "tilted" models where $\Omega = 1$, $\Lambda = 0$, but $n \leq 1$ varies.
3) $\Lambda$CDM, where $\Omega \leq 1$ for the matter, but it is compensated by a positive cosmological constant such that the universe remains flat as predicted by Inflation, $\Omega + \Lambda = 1$. We limit the discussion of this model to $n = 1$

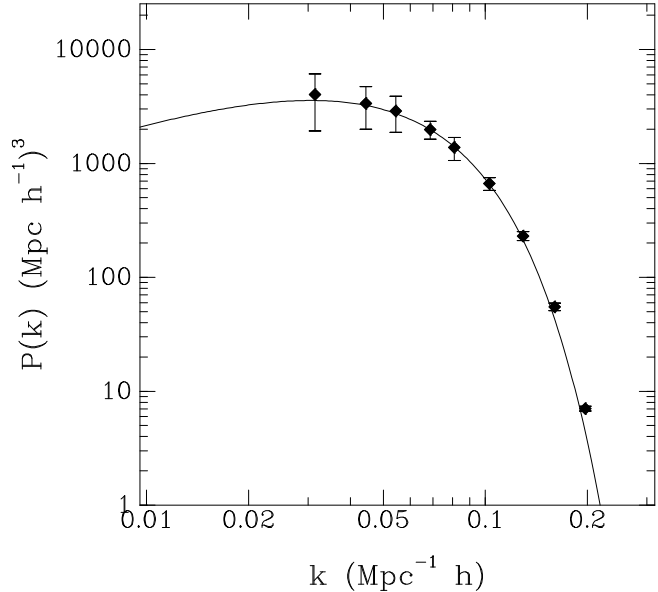

**Figure 7:** The nonlinear PS of sCDM simulations at $\sigma_8 = 0.77$, after G12 smoothing. The average and standard deviation of 20 simulations are shown in comparison with the G12-smoothed linear PS of $\sigma_8 = 0.77$.

4) OCDM, open models where $\Omega < 1$ while $\Lambda = 0$ and $n = 1$.

The dimensionless transfer function, $T(k)$, is approaching unity on large scales, and its shape, in particular the length scale of the peak that it induces on the PS, depends on the cosmological parameters listed above. For the CDM models we adopt the functional fit by Sugiyama (1995, and private communication):

$$T(k) = \frac{\ln(1 + 2.3q))}{2.34q} \left[1 + 3.89q + (16.1q)^2 + (5.46q)^3 + (6.71q)^4\right]^{-1/4}, \qquad (11a)$$

$$q = k \left[\Omega h \, \exp(-\Omega_b - h_{50}^{1/2} \Omega_b / \Omega) \, (\mathrm{h\,Mpc}^{-1})\right]^{-1}. \qquad (11b)$$

We also consider the family of mixed dark matter models [CHDM, with one species of neutrino ($N_\nu = 1$) and $\Omega_\nu \geq 0$ (Klypin *et al.* 1993)], and adopt the transfer function in tabular form as computed by Holtzman (1989, and private communication). When necessary we use linear interpolation between the transfer functions as computed for discrete choices of the parameters.

The last family of models we consider is the baryonic universe, with low $\Omega$ and isocurvature fluctuations (PIB, Peebles 1987; Cen, Ostriker & Peebles 1993). The model power spectrum in the range $k = 0.0005 - 0.05\,\mathrm{h\,Mpc}^{-1}$ was tabulated by Peebles (private communication). In the range $k > 0.05$, the logarithmic slope is assumed to be $-1$, and $n = 1$ at $k < 0.0005$.

### 5.2. COBE Normalization

We let the normalization constant, $A$, be determined by the second-year data of COBE's DMR experiment, subject to the other model parameters (Górski 1994, Górski *et al.* 1994, Sugiyama 1995, White & Bunn 1995). In the present application, COBE's data is represented by one parameter, the effective quadruple termed $Q_{rms-ps}$, which, for adiabatic fluctuations, $n = 1$ and $\Omega = 1$ is currently estimated to be $20\mu K$. For the model-dependent translation into $A$, we use functional fits (Zaroubi *et al.* 1996) to the results of White & Bunn (1995), which model separately the $\Omega$ dependence and the $n$ dependence via

$$A = A_1(\Omega) A_2(n), \tag{12}$$

as specified below.

This translation depends, in particular, on the relative contribution of tensor fluctuations of gravitational radiation (T), which, unlike the scalar part (S), are not relevant to structure formation. The tensorial contribution is assumed to be related to $n$, as in many models of Inflation (*e.g.* Turner 1993; Crittenden *et al.* 1993), via

$$T/S = 7(1 - n), \tag{13}$$

and the relevant $Q_{rms-ps}$ is reduced accordingly by $[S/(S+T)]^{1/2}$.

For $\Lambda$CDM and TCDM we use:

$$\begin{aligned} \log A_1(\Omega) = &\ 7.93 - 8.33\Omega + 21.31\Omega^2 - 29.67\Omega^3 + 10.65\Omega^4 \\ &+ 15.42\Omega^5 - 6.04\Omega^6 - 13.97\Omega^7 + 8.61\Omega^8 \end{aligned} \tag{14a}$$

$$\log A_2(n) = \begin{cases} -2.78 + 2.78n & (T = 0) \\ -4.50 + 4.50n & (T \neq 0) \end{cases} . \tag{14b}$$

The fit quoted here is for $h = 0.5$; the $h$ dependence is weak. [We note that the adopted $n$-dependence by White & Bunn (1995) is slightly different from the fit obtained by Bennett *et al.* (1994).]

For OCDM and no tensor fluctuations ($T = 0$) the fit is:

$$\log A_1(\Omega) = 5.75 + 1.68\Omega - 4.53\Omega^2 + 7.57\Omega^3 - 7.53\Omega^4 + 3.15\Omega^5 - 0.23\Omega^6 \tag{15a}$$

$$\log A_2(n) = -2.71 + 2.71n . \tag{15b}$$

For CHDM, as for sCDM, we use:

$$A = \frac{6\pi^2}{5} R_H^4 \left( \frac{Q_{rms-ps}}{T_0} \right)^2, \tag{16}$$

where $R_H = 6000\ h^{-1}$Mpc is the angular distance to the horizon, $T_0 = 2.726°K$ (Mather *et al.* 1994), and $Q_{rms-ps} = 20\mu K$.

When we need to normalize the PIB model we adopt $\sigma_8 = 1$, assuming no significant biasing for optical galaxies. This is roughly consistent with COBE's normalization.

### 5.3 Straightforward Comparison

Figure 8 compares the mass $P(k)f^2$ as derived from peculiar velocities with theoretical power spectra of representative models from the families described above, all COBE normalized. The error bars attached to the data are the measurement errors of Fig. 5, referring to the uncertainty in recovering the PS in our local neighborhood (mostly due to the scatter in the distance indicators). The error bar attached to the sCDM curve represent the *cosmic scatter* (CS) at that wavenumber, corresponding to the possible difference between the PS in our local neighborhood and the universal PS. The cosmic scatter is derived from mock catalogs drawn, without noise, from twenty COBE-normalized N-body simulations of random realizations of this model. This can serve as an approximation for the cosmic scatter in the other models considered here, assuming that the ratio of scatter to signal is roughly constant. The actual error can be approximated by adding in quadrature the measurement error and the cosmic scatter. In the following, unless stated otherwise, we assume $\Omega_b h^2 = 0.0125$ and no tensor fluctuations ($T = 0$).

The sCDM model is "rejected" by the most reliable data point ($k = 0.1$) only at the $\sim 1.5\sigma$ level if $h = 0.5$ ($2\sigma$ if $h = 0.7$). For the TCDM family, with $\Omega = 1$, the best fit seems to be $n \simeq 0.8$ for $h = 0.5$ ($n \simeq 0.6 - 0.7$ for $h = 0.7$). TCDM with tensor fluctuations prefers $n \simeq 0.9$ (0.8). For the $\Lambda$CDM family, with $n = 1$, best fit seems to be $\Omega \simeq 0.6$ (0.4). For OCDM, with $\Lambda = 0$ and $n = 1$, it becomes $\Omega \simeq 0.7$ (0.55).

The CDM models all seem to predict a PS with a slope not as steep as the measured PS. The curves that fit best the two data points at $k \leq 0.1$ slightly overestimate the power at the small-scale point ($k = 0.17$). The family of CHDM models, shown in Figure 8e, may help in this respect. Indeed, the PS *shape* seems most compatible with the data for $\Omega_\nu \simeq 0.3$. The model seems to slightly overestimate the amplitude, but only at the $\sim 1\sigma$ level. A slight tilt in $n$, and/or a small tensorial component, and/or a slight decrease in the observed $Q_{rms-\rm ps}$, can bring this model into perfect agreement with the data. A higher value of $h$ would require larger deviations from the "standard" values of the above parameters for a perfect fit.

Finally, Figure 8f shows the PIB family of models with $\Omega = 0.1$, all normalized such that $\sigma_8 = 1$. It is clear that this model severely underestimates the power, by an order of magnitude, reflecting the large $f^2(\Omega)$ factor between the power spectra of velocity and density (the latter enters via the $\sigma_8$ normalization). Low values of $\Omega$, at the level of $\sim 0.1$, are clearly ruled out by the data (same as in Figs. 8c and 8d).

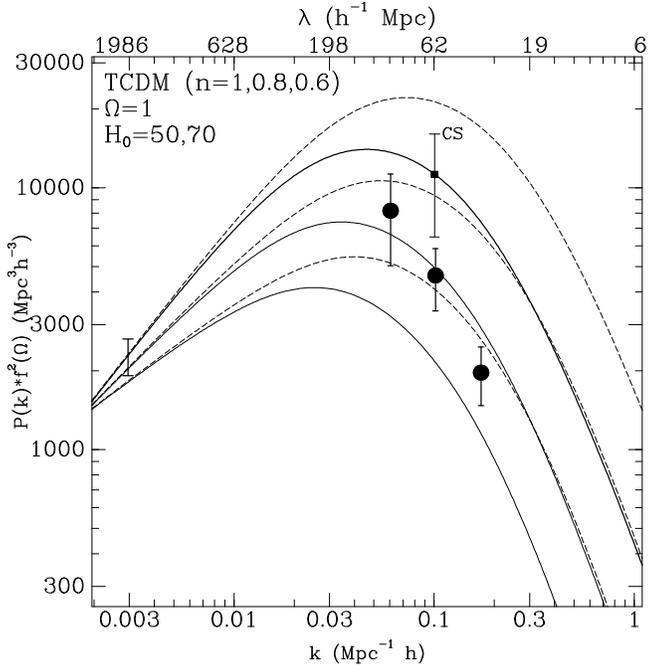
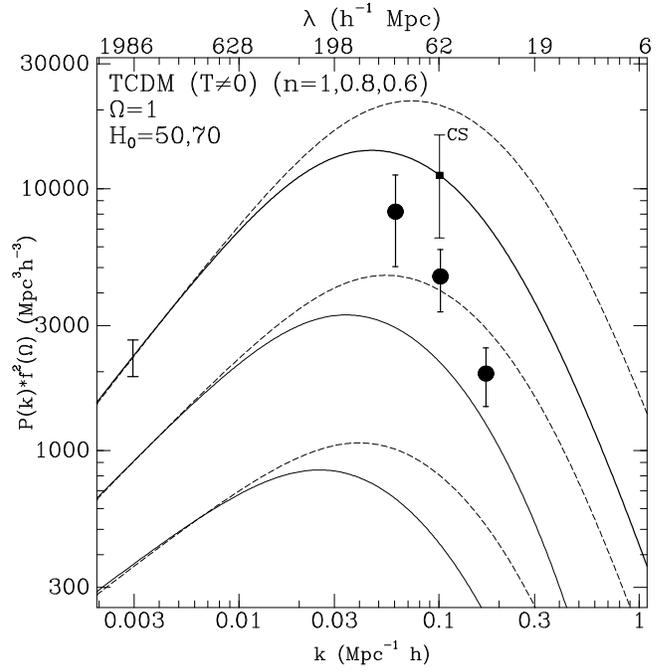
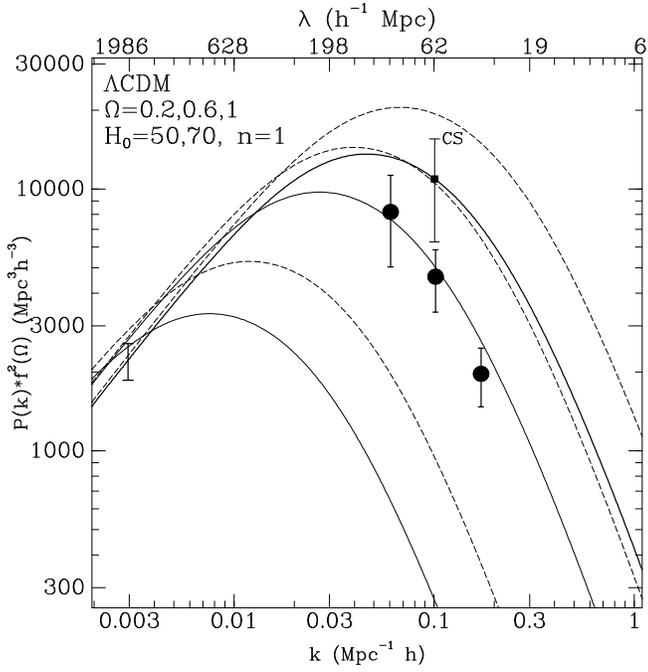
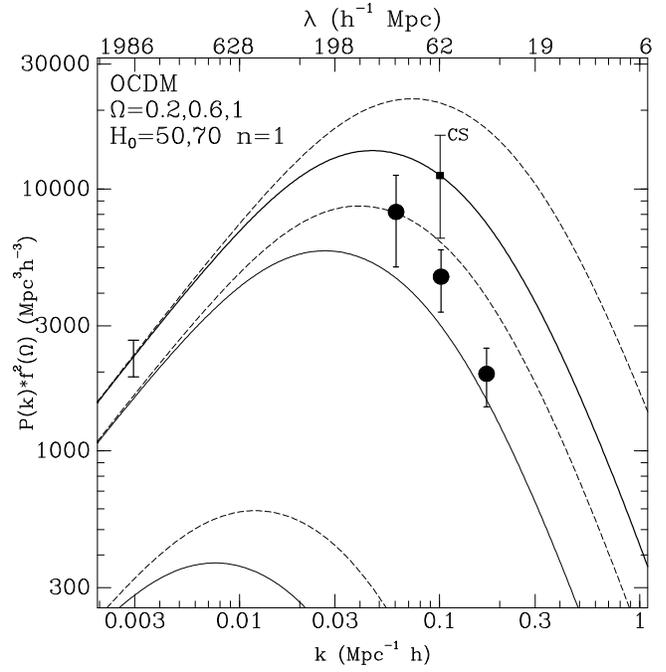

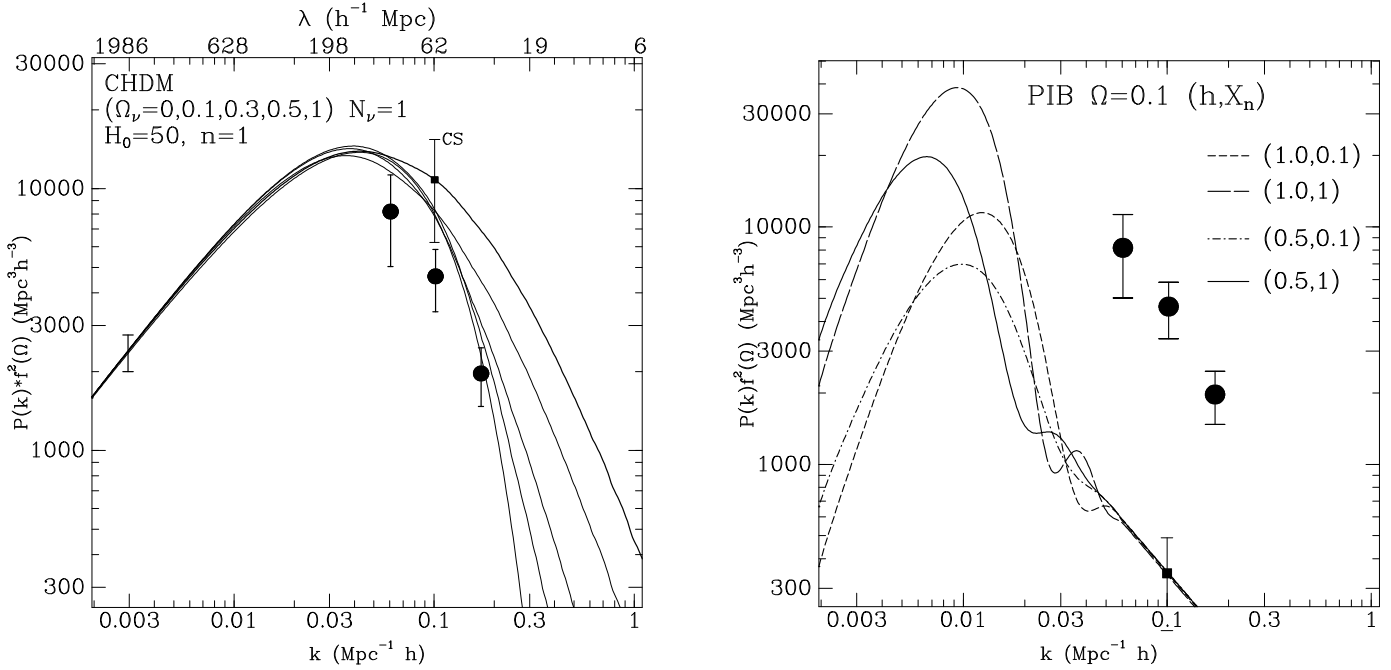

**Figure 8:** The mass density PS from peculiar velocities (filled circles and errors from Fig. 5) compared with theoretical models. The model power spectra are multiplied by $f^2(\Omega)$ to match the observed quantity. Cosmic scatter (CS) is computed from 20 sCDM simulations, and it should be added in quadrature to the measurement errors. The CDM models (a-e) are COBE normalized ($Q_{rms-\text{ps}} = 20\mu K$ for sCDM), and $\Omega_b = 0.0125\ h^{-2}$ is assumed. Solid and dashed curves correspond to $h = 0.5$ and 0.7 respectively. The sCDM model ($\Omega = 1$, $n = 1$, $h = 0.5$) is marked by a heavier line. Panels from upper left to bottom right, with the values of the variables corresponding to the curves from bottom to top: **(a)** TCDM $n = 0.6, 0.8, 1$. **(b)** TCDM with tensor fluctuations. **(c)** $\Lambda$CDM $\Omega = 0.2, 0.6, 1$. **(d)** OCDM $\Omega = 0.2, 0.6, 1$. **(e)** CHDM $\Omega_\nu = 0, 0.1, 0.3, 0.5, 1$. tabulated **(f)** PIB, with isocurvature fluctuations, $n = 1$, $\sigma_8 = 1$. The curves are characterized by $h$ and the ionization parameter $X_n$.

*5.4 Likelihood Analysis*

In order to further quantify the visual impressions from Figures 8, we perform an approximate likelihood analysis, by which we determine the best-fit parameters for each family of models, and estimate the goodness of fit.

For COBE-normalized models, we define a $\chi^2$ of the PS from velocities by

$$\chi_v^2 = \sum_{i=1}^{3} \frac{[P_{v\,obs}(k_i) - P_{v\,model}(k_i)]^2}{\sigma_v^2(k_i) + \sigma_{cs}^2(k_i)}\ , \qquad (17)$$

where $\sigma_v$ and $\sigma_{cs}$ are the measurement error and the estimated cosmic scatter respectively. Alternatively, we use $Q_{rms-\text{ps}}$ as an observed quantity ($Q_{obs}$), with a "measurement" error

$\sigma_Q$ of 8%, and add to the $\chi^2$ the term

$$\chi_Q^2 = \frac{(Q_{obs} - Q_{model})^2}{\sigma_Q^2} \ , \qquad (18)$$

where $Q_{model}$ is added as one of the free model parameters.

The sum in $\chi_v^2$ is over the three bins of $k$ at which the mass PS has been evaluated in §3. The choice of three bins is a compromise between the desire to extract as much information as possible from the given data and the wish to have the data points entering the sum as independent as possible from one another. The correlation between the bins is reduced by adopting bins of width corresponding to a wavelength range that is larger than the smoothing scale.

An approximate measure of goodness of fit is then provided by the incomplete gamma function

$$G = \Gamma(N_{dof}/2, \chi_{min}^2/2) \ , \qquad (19)$$

where $\chi_{min}^2$ is the value of $\chi^2$ at the best-fit parameters. The number of degrees of freedom, $N_{dof}$, for two free parameters in the model, is either one (when $\chi_v^2$ is minimized alone), or two (when $\chi_v^2 + \chi_Q^2$ is minimized). The quantity $G$ could be roughly interpreted as the probability that $\chi^2$ is as big as $\chi_{min}^2$ when the data and model are consistent with each other.

In order to test the independence of $P(k_i)$ in the three bins, the same $\chi^2$ analysis has been repeated using only the two disjoint bins at $k = 0.06$ and $0.17$, which are closer to being independent of each other, and alternatively using a single bin at $k = 0.1$, without changing the assumed error per bin. The goodness-of-fit for the two bins is found to be similar (within 20%) to the one obtained for the three bins. The minimum of $\chi^2$ for a single bin occurs at a similar point in parameter space (within 10%) to the minima obtained for three or two data points. These results are consistent with our earlier estimate of weak correlation between the bins based on the ratios of off-diagonal to diagonal terms in the covariance matrix (§3.2).

Figure 9 shows two-parameter contour maps of $\chi^2 - \chi_{min}^2$. In assigning probabilities to the contours, we make the simplifying assumption that this is a two-dimensional $\chi^2$ distribution with the two free parameters as variables. Then, The bold contours encompass the 68.3%, 95.4%, and 99.7% probabilities in the two-parameter plane. The error bars show the one-dimensional errors ($1\sigma$, $2\sigma$, $3\sigma$) in $\Omega$ for two different values of $h$, assuming that the latter are given *a priori*.

The conditional best-fit values of $\Omega$ can be summarized by the following, approximate, power-law relations, which roughly follow the $\chi^2$ "valleys" in the corresponding contour maps of Fig. 9, and the associated one-dimensional errors (the error variations can be read from the plots):

1) $\Lambda$CDM, $n = 1$: $\Omega \simeq 0.55 h_{50}^{-1.35} \pm 0.1$.
2) $\Lambda$/TCDM, $h = 0.5$: $\Omega \simeq 0.55 n^{-2.18} \pm 0.13$ ($T = 0$), $\Omega \simeq 0.57 n^{-4.32} \pm 0.10$ ($T \neq 0$).
3) OCDM, $n = 1$: $\Omega \simeq 0.67 h_{50}^{-0.75} \pm 0.07$.
4) O/TCDM, $h = 0.5$: $\Omega \simeq 0.5 n^{-1.27} \pm 0.13$ ($T = 0$).

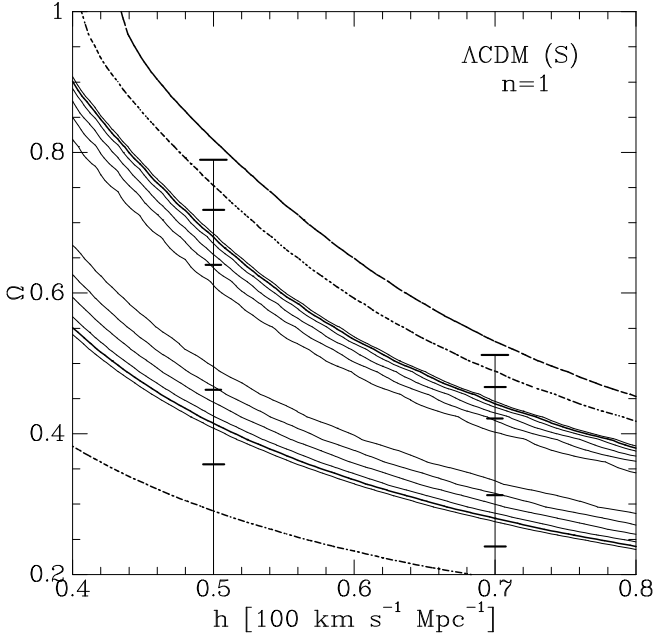
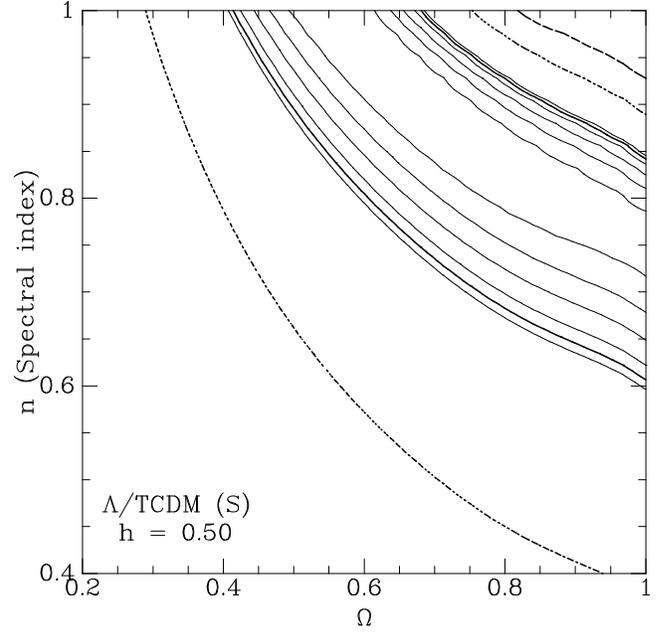
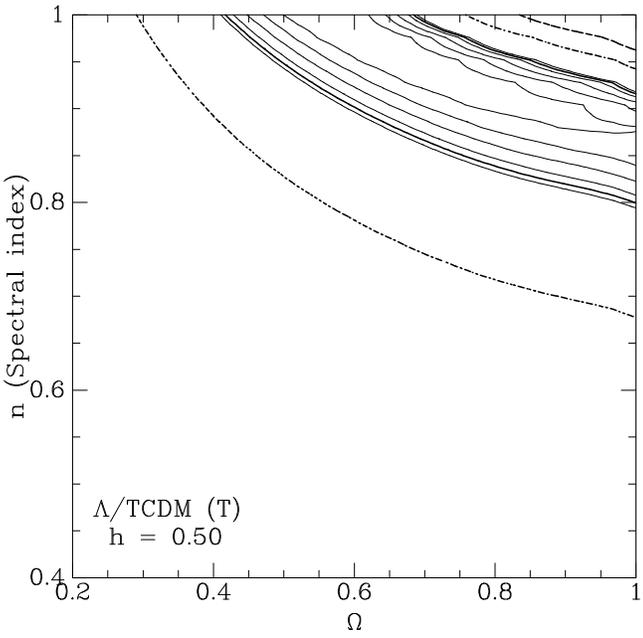
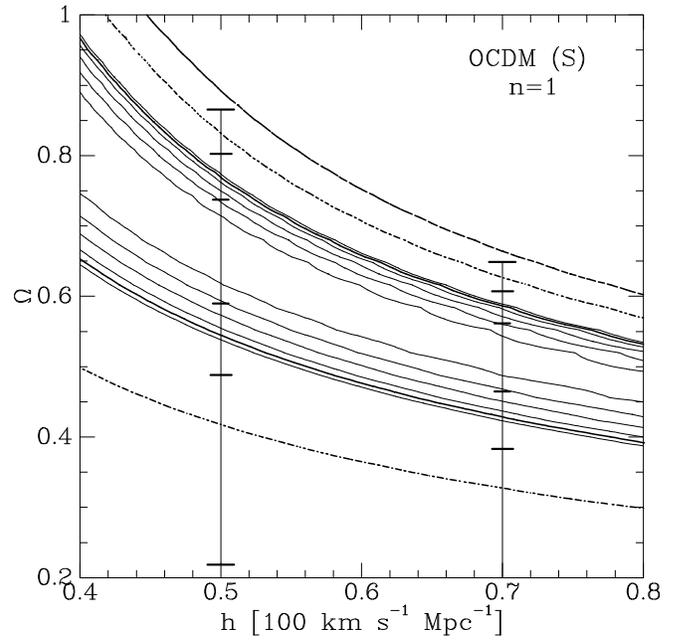

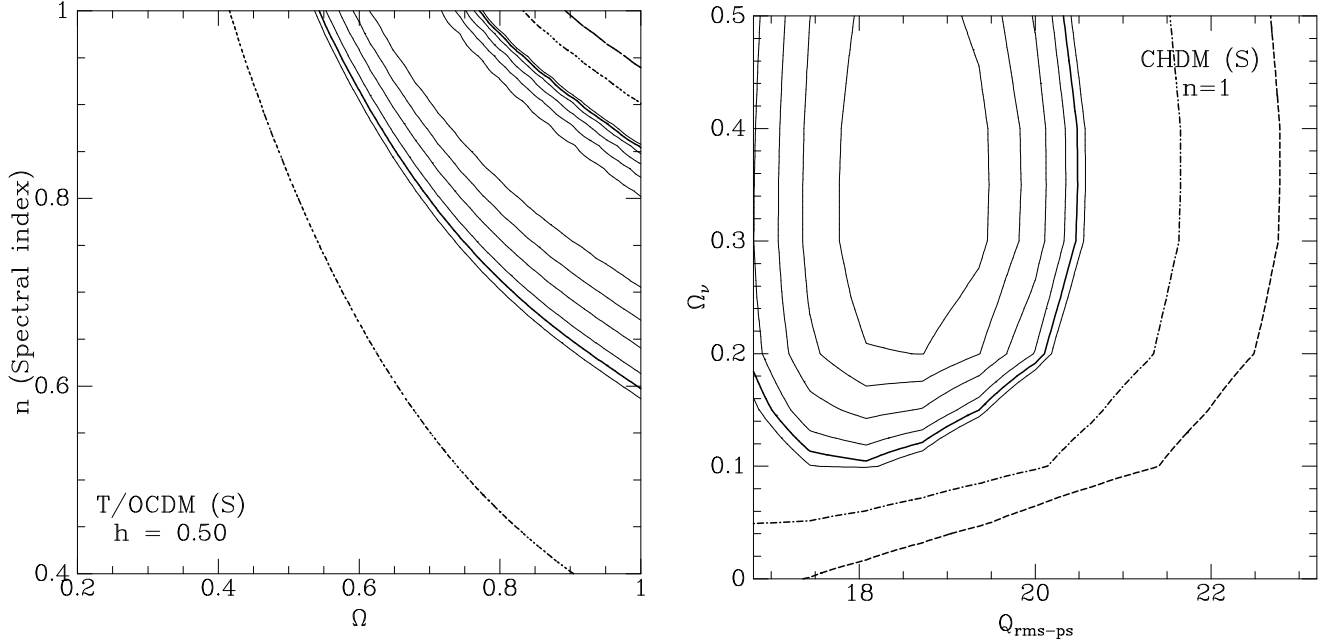

**Figure 9:** Contour maps of $\chi^2 - \chi^2_{min}$ for the different models. Spacing is $\Delta\chi^2 = 0.5$. Heavy contours mark the $1\sigma$, $2\sigma$ and $3\sigma$ confidence levels in the two-parameter plane (solid, short dash, and long dash). The error bars correspond to the conditional likelihood of the value of one parameter given the other fixed at a certain (arbitrary) value. The models are COBE normalized, except when $Q_{rms-ps}$ is taken to be one of the free parameters. The family of models and fixed parameters are specified in each panel. "S" and "T" stand for "scalar only" and "tensor+scalar" modes.

When we allow $Q_{rms-ps}$ to be one of the two free parameters, we obtain in the following models best fit at:

5) TCDM, $h = 0.5$, $T = 0$: $Q = 19.8\mu K$ and $n = 0.76$.

6) $\Lambda$CDM, $n = 1$, $h = 0.5$: $Q = 20.19\mu K$ and $\Omega = 0.55$.

7) OCDM, $n = 1$, $h = 0.5$: $Q = 19.80\mu K$ and $\Omega = 0.68$.

8) CHDM, $\Omega = 1$, $n = 1$, $h = 0.5$: $Q = 18.5\mu K$ and $\Omega_\nu = 0.36$.

Figure 10 shows the $\chi^2$ for the PIB model, fully ionized, with the normalization ($Q_{rms-ps}$, or $\sigma_8$) as a single free parameter. Best fit is found at $Q_{rms-ps} = 20.3\mu K$, which is very close to the value measured by COBE, and corresponding to a $\sigma_8$ value close to unity. The fit is dominated by COBE's normalization because of the small error associated with this measurement (8%) compared with the larger errors associated with the mass PS from peculiar velocities. The result is a significant deviation of the model from the velocity data, with a very poor goodness of fit.

Table 3 presents the approximate goodness of fit for the models we examined. For all the models except PIB, the fit, at the best-fit parameters, is acceptable. In the case of PIB, $G = 0.046$, meaning that the likelihood of the data given this model is extremely low. This is predominantly a reflection of the fact that the data disfavor very low values of $\Omega$, which are intrinsic to the PIB model.

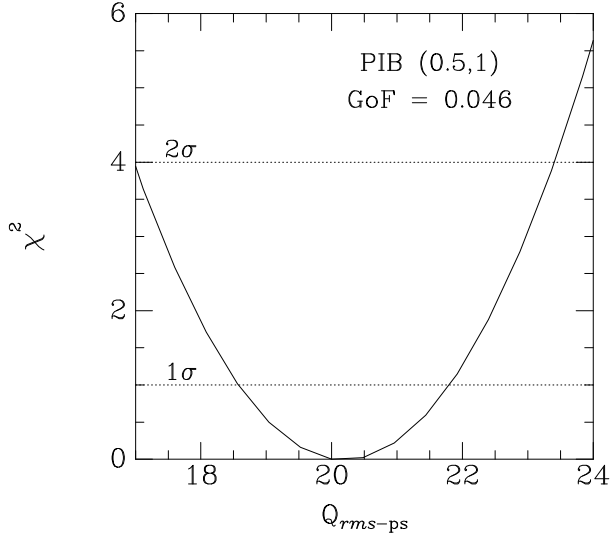

**Figure 10:** One-dimensional $\chi^2 - \chi^2_{min}$ for PIB ($\Omega = 0.1$, $h = 0.5$, fully ionized).

| Model | Variable 1 | Variable 2 | Fixed | GoF |
|---|---|---|---|---|
| $\Lambda$CDM | $h$ | $\Omega$ | $n=1$, $Q_{rms-ps}$ | 0.955 |
| $\Lambda$CDM | $Q_{rms-ps}$ | $\Omega$ | $h=0.5$, $n=1$ | 0.940 |
| $\Lambda$CDM | $\Omega$ | $n$ | $h=0.5$, $Q_{rms-ps}$, $T/S=0$ | 0.733 |
| $\Lambda$CDM | $\Omega$ | $n$ | $h=0.5$, $Q_{rms-ps}$, $T/S \neq 0$ | 0.712 |
| $\Lambda$CDM | $\Omega$ | $n$ | $h=0.75$, $Q_{rms-ps}$, $T/S=0$ | 0.748 |
| $\Lambda$CDM | $\Omega$ | $n$ | $h=0.75$, $Q_{rms-ps}$, $T/S \neq 0$ | 0.650 |
| CHDM | $Q_{rms-ps}$ | $\Omega_\nu$ | $h=0.5$, $\Omega=1$, $n=1$ | 0.339 |
| OCDM | $h$ | $\Omega$ | $n=1$, $Q_{rms-ps}$ | 0.624 |
| OCDM | $Q_{rms-ps}$ | $\Omega$ | $h=0.5$, $n=1$ | 0.858 |
| PIB | $Q_{rms-ps}$ | – | $h=0.5,1$, $\Omega=0.1$, $n=1$ | 0.046 |

**Table 3:** Approximate goodness of fit values for the best-fit models in each family.

Note that if we limit ourselves to modifying only one parameter of sCDM (*i.e.* changing $n$, or $\Omega$, or $\Omega_\nu$), we can reach almost perfect fit by a proper change in $n$ or $\Omega$. For a similar fit with CHDM, we need in addition a slight modification in either $n$ or $\Omega$, together with introducing $\Omega_\nu > 0$.

*5.5 Including Bulk Velocity*

The field of mass density fluctuations is computed from spatial derivatives of the peculiar velocity field, and it therefore contains only information on wavelengths smaller than the effective diameter of the volume sampled, $\sim 120$ h$^{-1}$Mpc. However, the velocity data contains additional information concerning density fluctuations on larger scales, e.g. via the bulk velocity over the whole observed volume. Dekel *et al.* (1996, in preparation; Dekel 1994) have computed the bulk velocity in a top-hat sphere of radius $R = 50$ h$^{-1}$Mpc about the Local Group from the same G12-smoothed velocity field recovered by POTENT from the Mark III data. They found $B_{obs} = 375 \pm 85$ km s$^{-1}$, with the error reflecting the standard deviation as computed from Monte Carlo mock catalogs plus systematic errors.

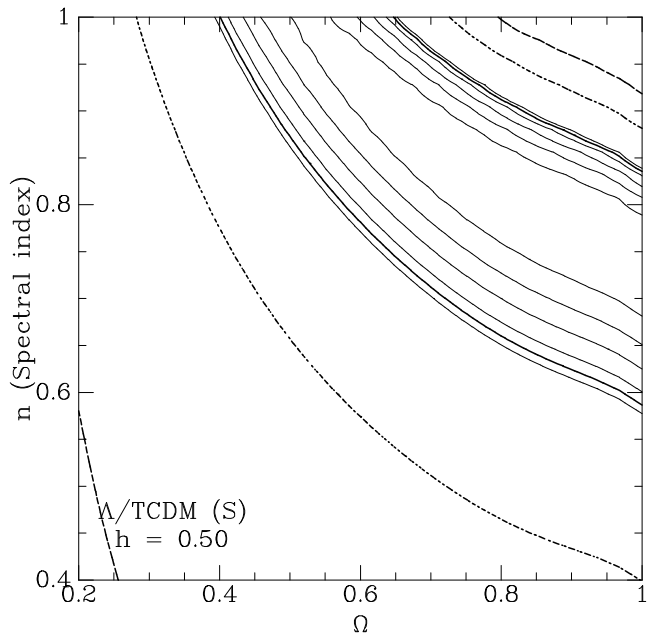

**Figure 11:** Contour map of $\chi^2 - \chi^2_{min}$ for T/$\Lambda$CDM, now using as additional data the bulk velocity in a top-hat sphere of radius 50 h$^{-1}$Mpc (compare to Fig. 9b).

The rms prediction of this quantity for each model is obtained by integrating the PS over large wavelengths,

$$B^2_{model} = \frac{f^2 H_0^2}{2\pi^2} \int_0^\infty dk\, P(k)\, \tilde{W}_R^2(k) \,, \qquad (20)$$

where $\tilde{W}_R(k)$ is the Fourier transform of the top-hat window of radius $R$. The cosmic scatter in this quantity for COBE-normalized sCDM is $\sigma_{cs}(B) \simeq 100$ km s$^{-1}$, and we adopt the same scatter-to-signal ratio for all other models.

When we wish to include the large-scale information of the velocity field, we add to the $\chi^2$ the term

$$\chi_B^2 = \frac{(B_{obs} - B_{model})^2}{\sigma^2(B) + \sigma_{cs}^2(B)} \,, \qquad (21)$$

assuming that $B$ is independent of the PS data in the three, smaller-scale bins. As an example of the effect of including the bulk velocity in the data, Figure 11 shows the revised $\chi^2$ map for the $\Lambda$CDM family of models with $\Omega$ and $n$ as free parameters and $h = 0.5$. A comparison with Fig. 9b indicates that the inclusion of the bulk velocity does not have a big effect: the overall change in the contour map is small, recovering a similar approximate relation of $\Omega \simeq 0.53 n^{-2.21} \pm 0.20$. The goodness of fit increases from $G = 0.733$ to $G = 0.748$.

# 6. CONCLUSION AND DISCUSSION

This is the first computation of the mass-density power spectrum from peculiar velocity data in a range corresponding to $30 - 120 \, h^{-1}$Mpc. The PS is based on the density field recovered by POTENT, within a sphere of effective radius $\simeq 60 \, h^{-1}$Mpc about the Local Group, from the Mark III Catalog of Peculiar Velocities.

A distinctive feature of the mass PS is its independence of any biasing of the galaxy density fields. We have assumed, however, that the galaxy velocities properly trace the underlying velocity field. The assertion that any velocity bias must be smaller than 20% is based on simulations (*e.g.* Carlberg 1993) and on the small variations between the velocity fields drawn from different galaxy types (Kolatt & Dekel 1994).

The method of computing the PS was calibrated and tested using detailed mock catalogs based on N-body simulations that closely mimic the real universe and the observational procedure. The errors in the resultant PS are quite big, but reasonably understood (for more details: Dekel 1994; Dekel *et al.* 1996; Dekel *et al.* 1996, in preparation) and properly estimated using Monte Carlo mock catalogs. This enables a quantitative comparison with other data and with theoretical models.

Our main straightforward result is that the mass PS, in the limited range $0.05 \leq k \leq 0.2 \, h \, \text{Mpc}^{-1}$ (on the short-wavelength side of the PS peak) can be approximated by the power law $P(k) = P_{0.1} \, (k/0.1 \, h \, \text{Mpc}^{-1})^{-m_{0.1}}$, with $m_{0.1} = 1.45 \pm \sim 0.50$, and $P_{0.1} = (4.6 \pm 1.4) \times 10^3 \, \Omega^{-1.2} \, (h^{-1}\text{Mpc})^3$ [or $\Delta^2_{0.1} = (0.233 \pm 0.071)\Omega^{-1.2}$].

Allowing for a turnover at $k_c$ into an asymptotic slope of $n = 1$ at $k \ll k_c$, the integral over the PS yields $\sigma_8 \Omega^{0.6} = 0.71 - 0.77 \, (\pm 0.12)$ for $k_c = 0.06 - 0.03$ respectively. This new result differs at about the $2\sigma$ level from the earlier estimate by Seljak & Bertschinger (1994) based on the Mark II sample, $\sigma_8 \Omega^{0.6} = 1.3 \pm 0.3$. The main improvements is that the current analysis includes five times denser sampling in an extended volume, better handling of systematic errors such as Malmquist bias, and better noise calibration based on detailed realistic mock catalogs. The current measurement can also be compared to the independent estimate based on cluster abundances, $\sigma_8 \Omega^{0.56} \simeq 0.57$ (White, Efstathiou, & Frenk 1993). The comparisons of the mass $\sigma_8$ to the values observed for optical galaxies ($\simeq 0.95$) and for IRAS 1.2Jy galaxies ($\simeq 0.6$) indicates $\beta \gtrsim 1$ for most galaxy types on these scales.

The mass PS in the observed range resembles in *shape* the PS of galaxies from angular surveys and redshift surveys. Direct comparisons of their amplitudes yield $\beta$ for the different galaxy surveys. Cosmic scatter is reduced because of partial overlaps between the volumes sampled by velocities and by galaxy density. We find that $\beta = (1.0 - 1.2) \pm 0.1$ for IRAS galaxies and $\beta = (0.8 - 1.0) \pm 0.1$ for optical galaxies. Thus, $\beta \simeq 1$ to within 30% for all the available galaxy types, indicating that $\Omega \simeq 1$ unless the galaxies are all antibiased. We attribute most of the variation in $\beta$ from survey to survey to the type dependence of the biasing parameter, but one must be cautioned that part of this variation may be attributed to errors in the galaxy power spectra (*e.g.* due to redshift distortions) and to non-trivial features in the biasing scheme.

Finally, we have combined the velocity PS with COBE's measurements to obtain constraints on the cosmological model and the initial fluctuations. We find that standard

CDM, with $\Omega = 1$ and $n = 1$, provides a relatively poor fit to the data, at the $2\sigma$ tail of the distribution (depending on $h$). One possibility that may help sCDM is that the COBE normalization adopted here is an overestimate, either because of a measurement error [e.g. Bennett et al. (1994) quote an amplitude lower by $\simeq 30\%$], or because of tensor fluctuations. Modifications of sCDM that provide better fits to the current data include: TCDM ($n \simeq 0.7$), $\Lambda$CDM ($\Omega \simeq 0.5$), OCDM ($\Omega \simeq 0.6$), and CHDM ($\Omega_\nu \simeq 0.3$). The mixed dark matter model, CHDM, provides the most appropriate PS shape in the relevant range of scales, but for best fit it needs a $10-20\%$ decrease in either $n$ or $Q_{rms-\mathrm{ps}}$ from their "standard" values (1 and $20\mu K$). Very open models with $\Omega \simeq 0.1 - 0.2$, and in particular the baryonic PIB model, are significantly ruled out.

What can we conclude from the various results summarized above about the value of $\Omega$ and the other theoretical ingredients involved? On one hand, the model-independent comparison of velocity and galaxy density PS yields $\beta \sim 1$, which indicates a high $\Omega$ (or, otherwise, severe anti-biasing). On the other hand, for $\Lambda$CDM and OCDM with $n = 1$, the comparison of the mass PS with COBE's measurements indicates lower values of $\Omega$, $\sim 0.5$, especially if $h$ is higher than 0.7. This discrepancy possibly indicates that the CDM transfer function with $n = 1$ is inappropriate — it does not have the necessary flexibility for a proper fit of the data without bias in the best-fit parameters. A high value of $\Omega$, consistent with the high value of $\beta$ obtained earlier, is easily achievable with an appropriate deviation from the $n = 1$ initial spectrum (as favored by Inflation models anyway), or with the inclusion of a hot dark matter component.

## Acknowledgment

We are grateful for close interaction with Saleem Zaroubi, for stimulating discussions with Joel Primack, and for very helpful comments by the referee, Michael Strauss. We thank the rest of the Mark III team, J. Willick, S.M Faber, S. Courteau and D. Burstein for allowing the use of preliminary versions of the Mark III catalog prior to publication. This work was supported in part by the US-Israel Binational Science Foundation grants 92-00355 and 95-00330, by the Israel Science Foundation grants 469/92 and 950/95, and by the US National Science Foundation grant PHY-91-06678.